# Uniform 2D Target Generation via Inverse-designed Metasurfaces

Yushi Zhou, Yun-Sheng Chen, *Member, IEEE*, and Yang Zhao, *Member, IEEE*

*Abstract*—We propose an inverse design framework for metasurfaces that achieves highly uniform two-dimensional intensity profiles across an on-demand shape. The optimization objective is formulated to enhance overall projection efficiency via the adjoint method, while a regularization term penalizes local deviations in field amplitude to suppress intensity non-uniformity. The regularization weight is adaptively tuned based on the current non-uniformity, enabling stable and efficient optimization. Compared with the widely used mean squared error (MSE) objective, our method yields superior performance in both intensity fidelity and uniformity. We also extend our framework to handle realistic Gaussian beam illumination by biasing the library. Simulation results confirm the effectiveness of our approach for generating high-quality, uniform field patterns.

## I. INTRODUCTION

Metasurfaces, as artificially engineered subwavelength structures, are highly effective in manipulating electromagnetic waves, enabling precise wavefront shaping[1], [2]. With the capability to control amplitude, phase, and polarization[3], [4] with high precision, metasurfaces have shown wide-ranging applications in various optical domains, including virtual-reality (VR) platforms[5], metasurface-driven OLEDs[6], holographs[7], and advanced beam shaping for integrated laser sources[8].

Metasurface design strategies can generally be categorized into forward and inverse design approaches. Forward design involves selecting the geometric parameters $(x_1, x_2, x_3, \ldots)$ of each meta-unit based on analytically or empirically derived electromagnetic responses, such as amplitude and phase. While this method is physically intuitive and offers insight into the underlying mechanisms, it requires manual tuning of the parameters for each individual meta-unit. As the size and complexity of the metasurface increase, this process becomes computationally intensive and ultimately impractical for large-scale implementations. Moreover, forward design is only capable of designing physically intuitive functions, such as beam focusing[9] and bending[10].

Inverse design has emerged as a powerful alternative that overcomes these limitations[11]. Starting from random initial parameters, inverse design frameworks iteratively optimize the geometric parameters of all meta-units simultaneously by optimizing a loss function that quantifies the deviation between simulated and target fields. This process, which is based on repeated forward simulations and backward updates, enables the efficient realization of complex metasurface functionalities. Inverse design has already demonstrated success in fabricating large area metalenses operating in the visible regime[12] and has shown superior performance in achieving multi-wavelength and multi-angle focusing with high efficiency[13].

However, despite these advancements, the majority of inverse-designed metasurfaces have focused on relatively simple target fields, primarily single focal points[13], [14], [15], while largely overlooking more complex field distributions. One particularly important target is the generation of uniform 2D intensity profiles. Such profiles are critical in applications requiring spatially homogeneous illumination, including flat-top laser beams for material processing[16] and sample excitation in wide-field microscopy[17]. For generating simpler patterns like square or circular flat-top beams, many existing studies have applied phase-retrieval strategies such as the Gerchberg-Saxton (GS) algorithm[18]. These iterative methods, however, often struggle to generate truly arbitrary, high-fidelity intensity profiles, particularly for non-convex shapes with sharp features. Furthermore, their convergence can be slow and highly sensitive to initialization[19], which can hinder design reliability and necessitate increased hardware complexity[16] to compensate for algorithmic constraints. For creating complex and arbitrary field distributions with high performance, direct, gradient-based inverse design has emerged as a more powerful and flexible paradigm, establishing a physical link between the device's geometric parameters and the desired optical performance.

The efficacy of any gradient-based optimization is critically determined by the formulation of its objective function, which defines the mathematical landscape the optimizer must navigate. A conventional and straightforward approach is to minimize the Mean Squared Error (MSE) between the simulated and target fields on a pixel-by-pixel basis, a method employed in various inverse design frameworks for holographic optimization[20], [21]. While intuitive, this pixel-wise objective function can create a highly non-convex optimization landscape. This makes the design process susceptible to being trapped in suboptimal local minima, often

(Corresponding author: Yang Zhao.)
Yushi Zhou, Yun-Sheng Chen, and Yang Zhao are with the Department of Electrical and Computer Engineering, Grainger College of Engineering, University of Illinois Urbana–Champaign, Urbana, IL 61801 USA (e-mail: yzhaoui@illinois.edu).

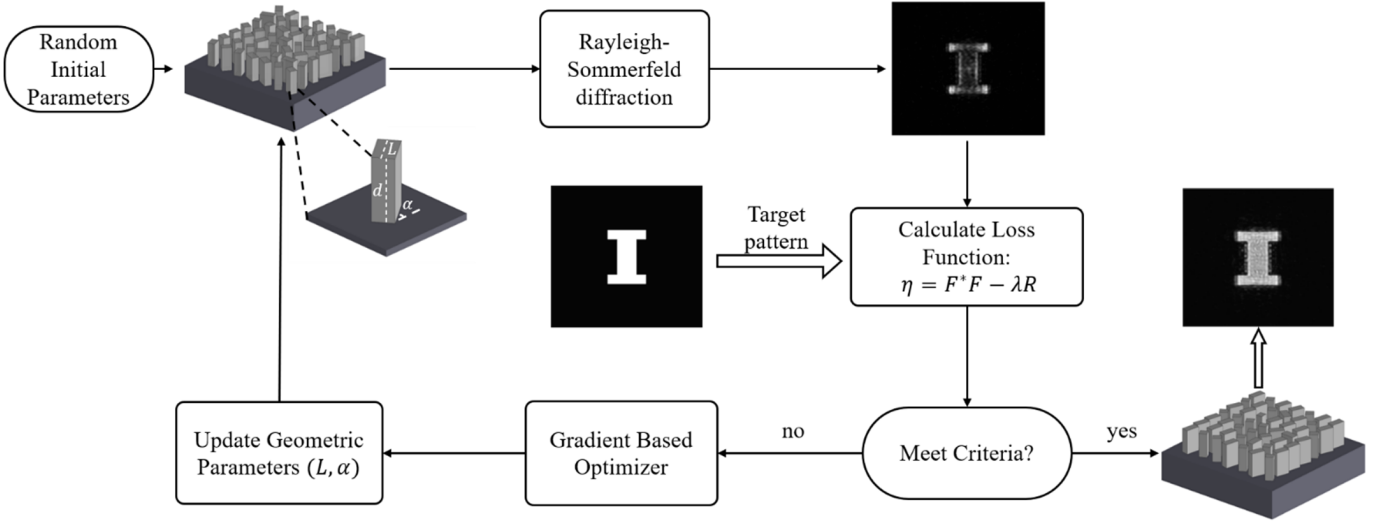

**Fig 1.** Flowchart of the inverse design process for a metasurface using the adjoint method, where the parameters $L$ and $\alpha$ are iteratively updated to obtain the optimized structure that produces the desired optical field.

sacrificing global performance metrics like overall efficiency and uniformity in favor of local, point-by-point accuracy. In this work, we propose and demonstrate an adjoint-based inverse design framework guided by a physically motivated objective function that overcomes these limitations. Our objective function is constructed to simultaneously maximize the total power projected into the target region, a global metric that promotes a smoother optimization landscape, while a dedicated regularization term directly penalizes intensity non-uniformity. This formulation allows our framework to significantly outperform conventional MSE-based inverse design, achieving substantially higher projection efficiency and superior field uniformity for arbitrary 2D target shapes.

## II. 2. FORWARD SIMULATION

A typical inverse design framework consists of two key components: (1) a fast and accurate forward simulation library that maps meta-unit geometric parameters to electromagnetic field responses, and (2) an inverse optimization algorithm that iteratively updates these geometric parameters to match a desired target field.

### A. Meta-unit Library

To achieve complete wavefront control, a full $2\pi$ phase modulation range is essential. However, for a single resonant antenna, the phase shift attainable by varying its length is typically limited to $\pi$[10]. To overcome this limitation, we employ the geometric phase mechanism, also known as the Pancharatnam-Berry (PB) phase[22]. This approach leverages the structural asymmetry of meta-units, which exhibit anisotropic refractive indices along two orthogonal directions. As a result, such a meta-unit can convert incident circularly polarized light into its opposite handedness. For example, when left circularly polarized (LCP) light is incident from the bottom of the meta-unit, the coefficient of the right circularly polarized (RCP) component emerging from the top surface is

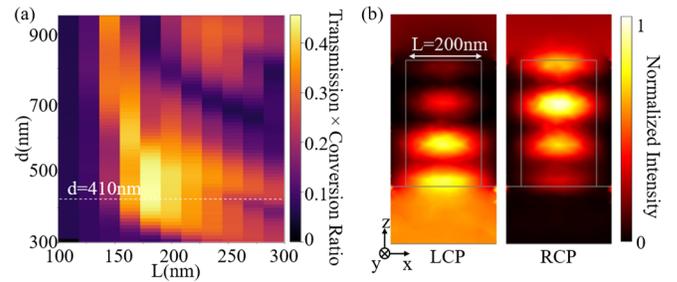

**Fig 2.** Simulation results of transmission and polarization conversion efficiency for rectangular meta-units under 800 nm wavelength illumination. (a) Heatmap of transmission × LCP-to-RCP conversion efficiency as a function of meta-unit height ($d$) and length ($L$), with fixed width $W = 100$ nm and period = 335 nm. The optimal region is observed near $d = 410$ nm and $L = 200$ nm. (b) Cross-sectional intensity profiles of LCP and RCP light when $L = 200$ nm, demonstrating strong transmission and efficient polarization conversion.

given by:

$$Ae^{i\psi} = i\sin\left(\frac{k_0 d\left(n_o - n_e\right)}{2}\right)$$
$$\times \exp\left(i\left(\frac{k_0 d\left(n_o + n_e\right)}{2} + 2\alpha\right)\right) \tag{1}$$

where $A$ and $\psi$ is the amplitude and phase of the RCP field, $k_0$ is the wave number, $d$ is the unit thickness, and $n_0$, $n_e$ are the ordinary and extraordinary refractive indices and $\alpha$ is the rotation angle, respectively[12].

The above expression suggests that when LCP light is incident, rotating a meta-unit by an angle $\alpha$ imparts a geometric phase shift of $2\alpha$ to the resulting RCP component. While this geometric phase mechanism allows full $2\pi$ phase control through rotation alone, such pure-phase designs often struggle to accurately reproduce features with high spatial frequencies, particularly near sharp edges, due to the lack of independent amplitude modulation. In contrast, phase-amplitude control offers greater design flexibility by enabling separate tuning of amplitude and phase. This can be achieved by fixing the meta-unit width and height, using the rotation angle $\alpha$ to modulate the phase $\psi$, and varying the meta-unit length $L$ to control the amplitude $A$ [12].

To determine the optimal meta-unit height for effective amplitude control of 800 nm wavelength input, we model the meta-units as amorphous silicon (a-Si), a material with high refractive index suitable for strong phase and amplitude modulation at this wavelength. We fix the width $W$ =100 nm, the unit cell period to 335 nm, and set $\alpha$ =0, then sweep the length $L$ from 100 nm to 300 nm. The incident field is LCP with an intensity of $10^4$ W/m$^2$. By scanning the meta-unit height, we identify geometries that maximize the product of total transmission and LCP-to-RCP conversion efficiency, while also providing a broad amplitude modulation range. As shown in Fig. 2(a), we select an operating contour near a height of $d$ =410 nm. Fig. 2(b) presents the cross-sectional field intensities of the LCP and RCP components for a single meta-unit, confirming high polarization conversion of 70%.

Directly solving Maxwell's equations using brute-force methods such as the finite element method (FEM) or finite-difference time-domain method (FDTD) provides accurate results but becomes prohibitively expensive for large-scale inverse design problems involving thousands of meta-units. To overcome this limitation, we adopt the library-based approach that constructs a surrogate model by precomputing and interpolating the electromagnetic response of individual meta-units.

Each meta-unit is simulated under LCP input and periodic boundary conditions, a common approximation that assumes geometric parameters vary slowly between adjacent meta-units. While this method reduces modeling accuracy for high numerical aperture (NA) designs[23], where adjacent meta-units exhibit abrupt parameter changes, it drastically lowers the computational cost compared to full-wave simulations of the entire metasurface.

We used the COMSOL Multiphysics® to perform the FEM simulation of the initial library[24]. We first densely sample the geometric parameter space by sweeping the meta-unit length $L$ from 100 nm to 300 nm in 5 nm steps and the rotation angle $\alpha$ from 0 to $\pi$ in increments of $\pi$/50. This generates a base grid of RCP amplitude $A(L,\alpha)$ and phase $\psi(L,\alpha)$ responses. Since $L$ and $\alpha$ have different units and physical meanings, we then perform interpolation in two stages: first, we unwrap and interpolate phase data along $\alpha$, then interpolate along $L$. To ensure smooth and stable interpolation, we use Chebyshev polynomial interpolation[14].

## B. Propagation method

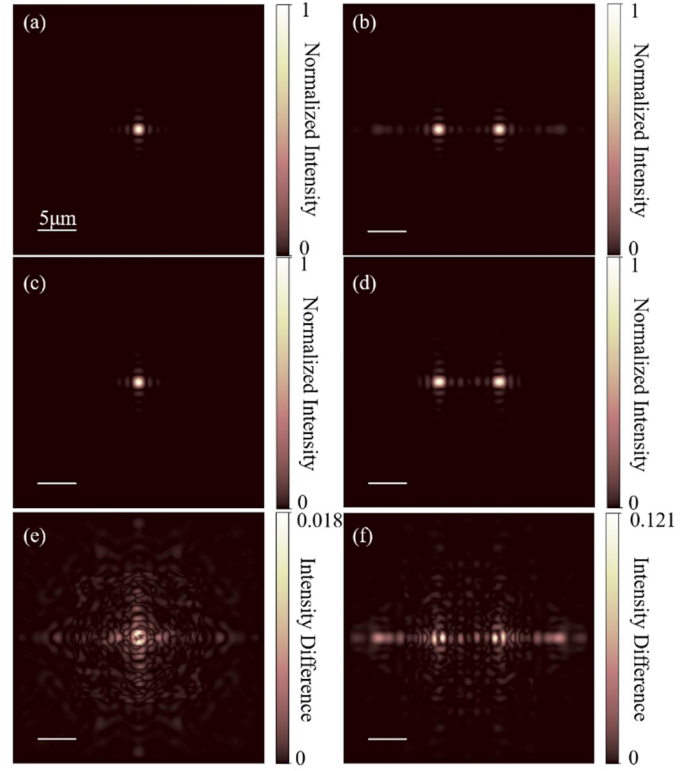

**Fig. 3.** Comparison of Rayleigh–Sommerfeld (R-S) and FDTD propagation results for single and dual-focus metalens designs. (a), (b) Intensity profiles at the focal plane calculated using R-S diffraction. (c), (d) Corresponding results from full-wave FDTD simulations (Tidy3D). (e), (f) Difference between R-S and FDTD results for each case. Scale bars: 5 μm.

To simulate the optical field propagation from the metasurface to the target plane, we employ the Rayleigh–Sommerfeld integral instead of the commonly used fast Fourier transform (FFT)-based methods. The Rayleigh–Sommerfeld approach is derived directly from the exact solution of the Helmholtz equation in free space and accounts for all diffraction orders. This makes it especially suitable for modeling higher NA metasurfaces, where paraxial approximations underlying FFT methods become inadequate. The Rayleigh-Sommerfeld diffraction is described as:

$$E(x,y,z) = \frac{1}{i\lambda} \iint_{-\infty}^{\infty} E(x',y',0) \frac{e^{ikr}}{r} \frac{z}{r}\left(1+\frac{i}{kr}\right)dx'dy' \quad (2)$$

where $E(x',y',0)$ denotes the complex field distribution at the metasurface plane ( $z = 0$), which is obtained via interpolation from the precomputed library. The variable $r$ is the distance between the observation point $(x,y,z)$ and the source point $(x',y',0)$, defined as:

$$r = \sqrt{(x-x')^2 + (y-y')^2 + z^2} \quad (3)$$

To validate the accuracy of the Rayleigh–Sommerfeld method, we compare its results against FDTD simulations using the Tidy3D solver[25]. Two test cases are considered: a single-focus metalens and a dual-focus metalens, each comprising a 50×50 array with a 20 μm-focal length.

Fig. 3 shows the comparison: the focal plane intensity profiles calculated using Rayleigh–Sommerfeld diffraction are shown in Fig. 3(a) and Fig. 3(b), while the corresponding FDTD results are in Fig. 3(c) and Fig. 3(d). The absolute difference between the two methods is visualized in Fig. 3(e) and Fig. 3(f).

To quantitatively assess the accuracy, we use the peak signal-to-noise ratio (PSNR), a common metric in image quality analysis, defined as[26]:

$$PSNR = 10 \cdot \log_{10}\left(\frac{MAX^2}{MSE}\right) \tag{4}$$

where MSE is the mean squared error over all pixels in the target region, and MAX is the maximum field intensity.

For the single-focus metalens, the PSNR between the Rayleigh–Sommerfeld and FDTD results is 65.02dB, indicating near-perfect agreement[27]. For the dual-focus design, the PSNR slightly decreases to 47.11dB, yet still reflects strong consistency. The minor degradation observed in more complex target patterns, despite having the same NA, is primarily due to increased local variation between adjacent meta-unit geometries. As the complexity of the target increases, the required spatial phase and amplitude distributions exhibit sharper gradients, causing neighboring meta-units to differ more significantly. This undermines the validity of the periodic approximation used in individual unit simulations, and the interpolated library becomes less accurate in predicting the true collective response of the metasurface under such rapidly varying configurations.

### III. 3.    Inverse design Methodology

#### A. Adjoint method for complex 2D target

For the reverse update of geometric parameters, deep learning has been shown to be an effective approach[28], [29]. However, neural networks heavily depend on the dataset, and their performance decreases for data points far from the training set. Moreover, acquiring the training dataset requires substantial computational resources. In contrast, the adjoint method offers a more physically interpretable and computationally efficient alternative. Unlike traditional brute-force approaches, the adjoint method updates the geometric parameters of all units on the metasurface with just one forward simulation and one adjoint simulation. The core principles of the adjoint method lie in Lorentz reciprocity and the symmetry of the Green's function[30]. Specifically, on the target image plane $S$, we define the projection of the field $\vec{E}$ into the target field $\overrightarrow{E_t}$ by the cross Poynting vector as:

$$F = \frac{1}{4P}\int_S\left(\vec{E}\times\overrightarrow{H_t^*} + \overrightarrow{E_t^*}\times\vec{H}\right)\cdot\vec{n}\,da$$
$$= \frac{1}{4P}\int_S\left[-\left(\vec{n}\times\overrightarrow{H_t^*}\right)\cdot\vec{E} + \left(\vec{n}\times\overrightarrow{E_t^*}\right)\cdot\vec{H}\right]da \tag{5}$$

where $P$ is the normalization power. The efficiency is then defined as:

$$\eta = F^* \cdot F \tag{6}$$

Based on the equivalence theorem[31], the equivalent electric and magnetic currents on the target image surface are defined as: $\overrightarrow{J_a} = -\vec{n}\times\overrightarrow{H_t^*}$ and $\overrightarrow{K_a} = -\vec{n}\times\overrightarrow{E_t^*}$. Substituting these relations, we have:

$$F = \frac{1}{4P}\int_S\left[\overrightarrow{J_a}\cdot\vec{E} - \overrightarrow{K_a}\cdot\vec{H}\right]da \tag{7}$$

While the equivalent sources only generate electromagnetic fields on one side of the boundary, which we define as the focal plane, the field on the opposite side is zero[32]. When a mirror-symmetric current distribution is subtracted, the magnetic current (a pseudovector) effectively doubles, while the electric current cancels out, leaving the field below the focal plane unchanged. Using the Lorentz reciprocity theorem:

$$\int_S\left(-\overrightarrow{K_a}\cdot\vec{H}\right)da = \int_M\left(-\vec{K}\cdot\overrightarrow{H_a}\right)da_M. \tag{8}$$

so

$$F = \frac{1}{4P}\int_M\left(-\vec{K}\cdot\overrightarrow{H_a}\right)da_M. \tag{9}$$

This restricts the integral region from free space to the metasurface region $M$, requiring only the calculation of the local equivalent current $\vec{K} = -\vec{n}\times\vec{E}$ induced by the metasurface and the adjoint field $\overrightarrow{H_a}$ from the target. Using Green's function, the adjoint field is expressed as:

$$\overrightarrow{H_a} = -\int_S\overrightarrow{K_a}(\vec{x})\cdot\frac{\partial G\left(\overrightarrow{x'},\vec{x}\right)}{\partial n}da \tag{10}$$

Given that the target amplitude distribution is uniform, the 2D problem can be simplified by discretizing the image into numerous "0D pixels" and then the normalized adjoint source becomes:

$$\left|\overrightarrow{K_a}\right| = \begin{cases} 1 \text{ in target pixel region } S_T \\ 0 \text{ in other pixel regions} \end{cases} \tag{11}$$

With the symmetry of the Green's function, $G\left(\vec{x},\overrightarrow{x'}\right) = G\left(\overrightarrow{x'},\vec{x}\right)$, the projection $F$ is given as:

$$F = \frac{1}{4P}\int_M\left(\vec{K}\left(\overrightarrow{x'}\right)\cdot\int_S\overrightarrow{K_a}(\vec{x})\cdot\frac{\partial G\left(\overrightarrow{x'},\vec{x}\right)}{\partial n}da\right)da_M$$
$$= \frac{1}{4P}\int_{S_T}\int_M\left(E\left(\overrightarrow{x'}\right)\cdot\frac{\partial G\left(\vec{x},\overrightarrow{x'}\right)}{\partial n'}\right)da_M\,da \tag{12}$$

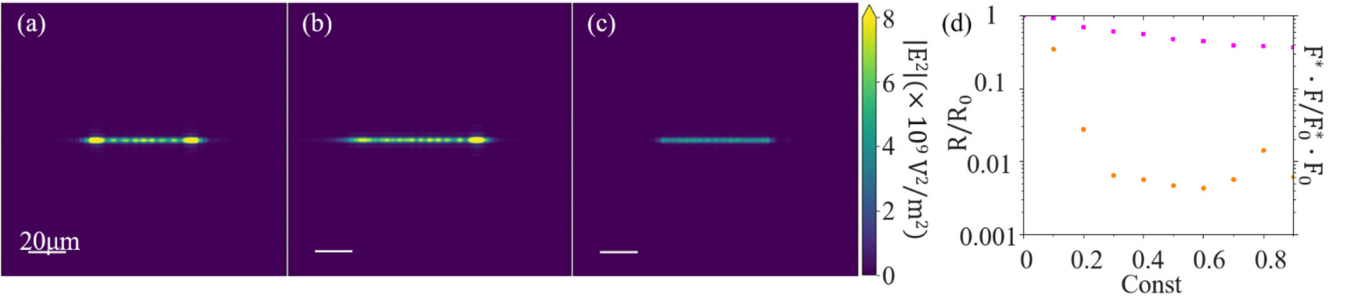

**Fig. 4.** Effect of regularization on field uniformity. (a) Optimized intensity pattern for a line-shaped target without regularization. (b), (c) Optimized results with regularization constants *const* = 0.1 and *const* = 0.3, respectively, showing progressively improved field uniformity. (d) Comparison of final efficiency $F^* \cdot F$ (circle) and regularization term $R$ (square) as functions of the regularization constant. Scale bars: 20 μm.

Here, $\int_M \left( \vec{E}(\vec{x}) \cdot \frac{\partial \vec{G}(\vec{x}, \vec{x'})}{\partial n'} \right) da_M$ is identical to (2), which gives the electric field at the target pixel. Maximizing the projection is equivalent to maximizing the sum of the electric field in the target region pixels, which aligns with the physical scenario. For a PB-phase metasurface composed of N units, the geometric parameters are represented as $\vec{p} = (L_1, L_2, \ldots, L_N, \alpha_1, \alpha_2, \ldots, \alpha_N)$. The gradient of $F$ at the target plane is calculated:

$$\nabla_{\vec{p}} F = \frac{1}{4P} \int_{S_T} \int_M \left( \nabla_{\vec{p}} \vec{E}(\vec{x'}) \cdot \frac{\partial \vec{G}(\vec{x}, \vec{x'})}{\partial n'} \right) da_M \, da \quad (13)$$

Green's functions are independent of geometric parameters and only need to be calculated once during the entire optimization. Using the gradient descent method, the parameters can be updated. In this work, the optimization was performed using the gradient-based method-of-moving-asymptotes (MMA) algorithm[33] and the NLopt library[34].

### B. Regularization for uniformity

The adjoint-based optimization method described above primarily aims to maximize the total intensity within the target region. However, this objective often results in highly non-uniform field distributions, where intensity is disproportionately concentrated near the edges or corners of the target. Such non-uniformities significantly diminish the quality and practical utility of the generated patterns, particularly in applications that demand uniform illumination. This issue is evident in Figs. 4(a) and 5(a), which show the results for a line-shaped and an "I"-shaped target, respectively. In both cases, the intensity distributions exhibit strong boundary accumulation, leading to a noticeable loss in pattern fidelity. In Figs. 4 and 5, the metasurface aperture has a focal length of 300 μm. The displayed intensity corresponds to RCP illumination and is normalized across the target region in each subplot.

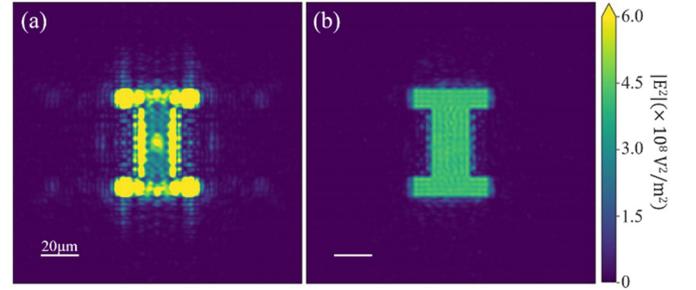

**Fig. 5.** (a) Optimized intensity pattern for an "I"-shaped target without regularization. (b) Improved intensity uniformity for the "I"-shaped target with optimal regularization. Scale bars: 20 μm.

To overcome this challenge, we introduce a regularization term into the adjoint optimization framework that penalizes deviations from a uniform field distribution across the target plane. The updated optimization objective function combines the original efficiency term with the regularization penalty:

$$\eta = F^* \cdot F - \lambda R \quad (14)$$

Here, $F^* \cdot F$ represents the conjugate square of the projection of the simulated field onto the target field (as derived in (6)), and $R$ is the penalty term representing the non-uniformity. The coefficient $\lambda$ adaptively controls the trade-off between maximizing efficiency and minimizing field variance inside the target area.

The regularization term $R$ is defined as the variance of the electric field amplitude across all N pixels in the target region, which is a measurement of non-uniformity:

$$R = \frac{1}{N} \sum_{i=1}^{N} \left( |E_i(\vec{x})| - \mu \right)^2 \quad (15)$$

where $|E_i(\vec{x})|$ is the electric field at pixel $i$ on the target plane, and $\mu$ is the mean amplitude across all target pixels: $\mu = \frac{1}{N} \sum_{i=1}^{N} |E_i(\vec{x})|$. This term encourages all field magnitudes to remain close to the average and promotes a more uniform intensity distribution.

To incorporate this regularization in a gradient-based optimization, we compute the derivative of $R$ with respect to

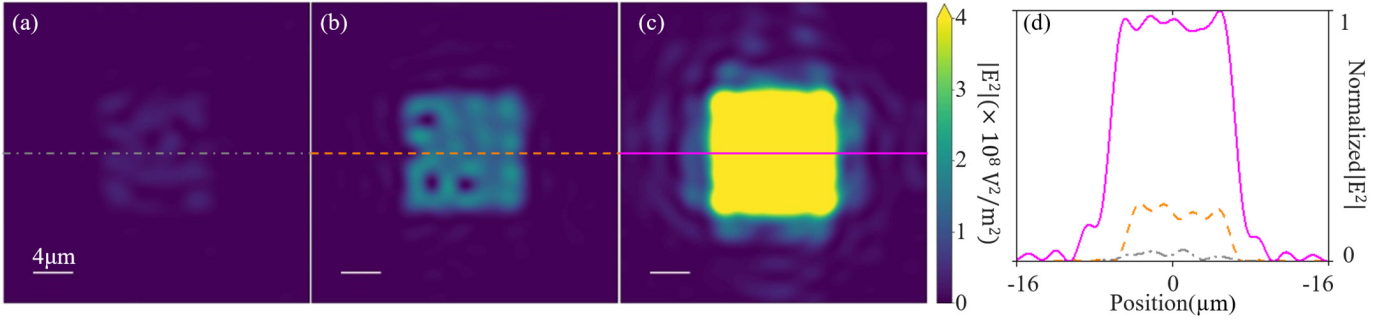

**Fig. 6.** Comparison of optimization results for square target images using different objective functions and optimizers. (a) Result obtained using MSE loss with MMA. (b) Result obtained using MSE loss with Adam. (c) Result obtained using the proposed loss with uniformity regularization ( *const* = 0.3) and MMA. (d) Normalized intensity line profiles along the dashed white line in (a–c), showing the line-profile performance of each case (a, gray dash-dot; b, orange dash; c, pink solid lines). The metasurface consists of 100 × 100 units with a focal length of 100 μm. All intensity distributions correspond to normalized RCP illumination. Scale bars: 4 μm.

the complex field value $E_j$. Applying the chain rule, we obtain:

$$\frac{\partial R}{\partial E_j} = \frac{2}{N}\left(\left|E_j\right| - \mu\right)\left(\frac{\partial \left|E_j\right|}{\partial E_j} - \frac{1}{N}\frac{\partial \left|E_j\right|}{\partial E_j}\right)$$
$$- 2\frac{\partial \left|E_j\right|}{\partial E_j}\frac{1}{N^2}\sum_{i \neq j}\left(\left|E_i\right| - \mu\right)$$

(16)

Using the identity:

$$\sum_{i=1}^{N}\left(\left|E_i\right| - \mu\right) = 0 \Rightarrow \sum_{i \neq j}\left(\left|E_i\right| - \mu\right) = -\left(\left|E_j\right| - \mu\right)$$

(17)

and the derivative:

$$\frac{\partial \left|E_j\right|}{\partial E_j} = \frac{1}{2\left|E_j\right|}\overline{E_j}$$

(18)

we arrive at the simplified and computationally efficient expression:

$$\frac{\partial R}{\partial E_j} = \frac{\left(\left|E_j\right| - \mu\right)}{N} \cdot \frac{\overline{E_j}}{\left|E_j\right|}$$

(19)

This form enables end-to-end backpropagation of the regularization gradient to the metasurface parameters through the adjoint method. To ensure the regularization contributes meaningfully during each iteration, we define the regularization weight $\lambda$ adaptively:

$$\lambda = const \cdot \frac{\left|F^* \cdot F\right|}{R + \epsilon}$$

(20)

Here, *const* is a tunable hyperparameter (typically set between 0.1 and 0.5) set for one optimization, and $\epsilon = 10^{-9}$ is a small constant to prevent singularity. This adaptive weighting in each iteration ensures that the impact of the regularization term is relatively larger when the field non-uniformity is high and diminishes when the distribution becomes more uniform.

By introducing this regularization term, we effectively suppress intensity peaks and achieve significantly improved field uniformity, as demonstrated in Figs. 4(b) and 4(c). These results correspond to the metasurface optimized for a line target using regularization constants *const* = 0.1 and 0.3, respectively, showing progressive smoothing of intensity variations.

The influence of the regularization strength is quantitatively summarized in Fig. 4(d), which plots both the final optimized $F^* \cdot F$ efficiency and regularization term $R$ of the line target as functions of the regularization constant, the values of both terms are normalized by the relative values without regularization. Notably, the non-uniformity $R$ drops significantly when *const* increases to around 0.2–0.3, while the efficiency $F^* \cdot F$ slightly decreases but remains within the same order of magnitude. As shown in Fig. 5(b), for the "I"-shaped target, applying regularization significantly enhances the pattern uniformity compared to the unregularized result in Fig. 5(a).

### C. Comparison with MSE objective function

An alternative approach to constructing the loss function for inverse design via gradient descent is to minimize the MSE between the actual field and the target field at each pixel. The MSE objective function is given by:

$$l = \frac{1}{N}\sum_{i=1}^{N}\left(\left|E\left(x_i, y_i, z\right)\right|^2 - \left|E_{tar}\left(x_i, y_i, z\right)\right|^2\right)^2$$

(21)

This pixel-wise formulation has been employed in both brute-force machine learning backpropagation frameworks[21] and in end-to-end physically interpretable adjoint-based methods[20] to perform holographic optimization.

However, compared to our proposed objective function in (7), which maximizes the total intensity within the target region and thereby defines a smoother, more convex optimization landscape, the conventional MSE objective is significantly more prone to local optima. This limitation becomes apparent in the design of a 100 × 100 meta-unit metasurface targeting a square intensity distribution. Pixel-wise squared-error minimization creates a highly rugged loss surface, which poses difficulties for the MMA method, as shown in Fig. 6(a), resulting in poor efficiency ($F^*F = 1.38 \times 10^9$). The Adam optimizer, which incorporates momentum and RMSprop to better navigate non-convex spaces, produced improved performance ($F^*F = 6.16 \times 10^{10}$), as illustrated in Fig. 6(b). In contrast, our adjoint-based method with the proposed loss achieved an efficiency of $F^*F = 1.96 \times 10^{12}$ —over two orders of magnitude higher than MSE+MMA and one order higher than MSE+Adam—while simultaneously yielding the most uniform field ($R = 1.47 \times 10^6$). The corresponding MSE value ($3.63 \times 10^{-5}$) remains comparable to MSE+Adam ($2.7 \times 10^{-5}$) and substantially smaller than MSE+MMA ($2.48 \times 10^{-4}$). As further confirmed by the line profiles in Fig. 6(d), our formulation delivers both higher efficiency and improved field uniformity without compromising pixel-wise accuracy.

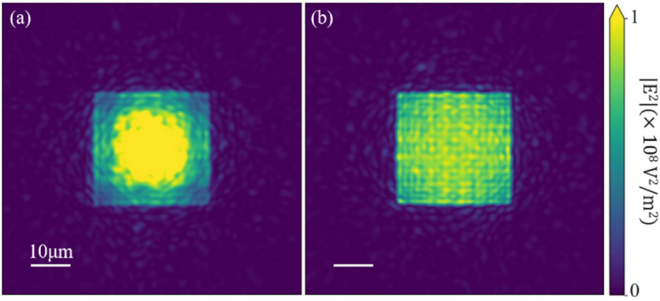

**Fig. 7.** Impact of Gaussian beam illumination and correction via input-aware optimization. (a) A metasurface ($100 \times 100$ units, ~30 µm) designed for plane-wave produces a distorted square target when illuminated by a Gaussian beam (FWHM = 100 µm, focus = 50 µm). (b) Re-optimized design using a Gaussian-weighted library restores the square shape under the same input. Scale bars: 10 µm.

### IV. 4. Gaussian input

All previous simulations were conducted using a metasurface library constructed under the assumption of a normally incident plane wave. However, practical laser sources typically exhibit a Gaussian intensity profile:

$$I(r) = I_0 \exp\left(-\frac{2r^2}{w_0^2}\right) \tag{22}$$

where r is the radial distance from the beam center, and $w_0$ is the beam waist. When a metasurface designed for plane-wave illumination is used under Gaussian beam excitation, its performance can degrade significantly. This phenomenon is illustrated in Fig. 7(a), where a metasurface consisting of $100 \times 100$ units (approximate size: 30 µm) is designed to focus on a 50 µm focus plane square target. The Gaussian beam has a full width at half maximum (FWHM) of 100 µm. Under such conditions, the resulting intensity profile at the target plane no longer preserves the uniform square shape.

To address this mismatch, we leverage the flexibility of the amplitude. Specifically, we introduce a spatial bias into the library by re-weighting the amplitude contribution of each meta-unit according to the Gaussian profile of the actual input beam. The central meta-units are assigned higher amplitudes consistent with the Gaussian envelope:

$$\sigma = \frac{\text{FWHM}}{2\sqrt{2\ln 2}} \tag{23}$$

$$\text{intensity\_factor}(r) = \exp\left(-\frac{r^2}{2\sigma^2}\right) \tag{24}$$

By integrating this intensity factor into the library construction and performing re-optimization accordingly, the resulting metasurface better accommodates the actual input profile. The improved performance is shown in Fig. 7(b), where the reconstructed square target is significantly more accurate under the same Gaussian beam illumination.

### V. Conclusion

We have presented a comprehensive inverse design framework for metasurfaces that enables the generation of uniform intensity profiles across arbitrary target shapes. By introducing a physically motivated regularization term into the adjoint-based optimization process, we suppress local intensity peaks and improve pattern fidelity. Comparative experiments demonstrate that our objective function outperforms conventional MSE-based designs in both stability and uniformity. We also extend the approach to account for realistic Gaussian beam illumination by introducing a spatial bias in the metasurface library. While this study uses a library-based approach for single-wavelength operation, the framework is generalizable and can be extended to multi-wavelength targets or full-wave simulations when sufficient computational resources are available.

**Yushi Zhou** received the B.S. degree in electrical engineering from Tsinghua University, Beijing, China. He is currently working toward the Ph.D. degree with the Department of Electrical and Computer Engineering, University of Illinois Urbana–Champaign, Urbana, IL, USA. His research interests include nanophotonics, metasurfaces, and inverse design.

**Yun-Sheng Chen** received the M.S. degrees in materials science and engineering from National Sun Yat-Sen University, Kaohsiung, Taiwan, and in optics from CREOL, University of Central Florida, Orlando, FL, USA, and the Ph.D. degree in electrical engineering from the University of Texas at Austin, Austin, TX, USA. He was a Postdoctoral Fellow with the Molecular Imaging Program at Stanford (MIPS), Stanford University, Stanford, CA, USA. He is currently an Assistant Professor with the Department of Electrical and Computer Engineering, University of Illinois Urbana–Champaign, Urbana, IL, USA. His research interests include imaging, diagnostic, and therapeutic techniques using light and ultrasound.

**Yang Zhao** received the Ph.D. degree in electrical and computer engineering from the University of Texas at Austin, Austin, TX, USA. She was a Postdoctoral Fellow with the Department of Materials Science, Stanford University, Stanford, CA, USA, before joining the University of Illinois Urbana–Champaign, Urbana, IL, USA. She is currently an Assistant Professor with the Department of Electrical and Computer Engineering at UIUC.